# Relationship between polymorphic structures and magnetic properties of La$_{2-x}$A'$_x$Ni$_7$ compounds (A' = Sm, Gd)

Valérie Paul-Boncour[1*], Véronique Charbonnier[1,2], Nicolas Madern[1,3], Lotfi Bessais[1], Judith Monnier[1], Junxian Zhang[1]

[1]Univ. Paris-Est Créteil, CNRS, ICMPE, UMR7182, F-94320 Thiais, France
[2]Energy Process Research Institute, National Institute of Advanced Industrial Science and Technology (AIST), Tsukuba West, 16-1 Onogawa, Tsukuba, Ibaraki 305-8569, Japan.
[3] Mines Paris, PSL University, MAT - Centre des Matériaux, CNRS UMR 7633, BP 87, 91003, Evry, France

**Abstract**

$A_2$Ni$_7$ compounds ($A$ = rare earth) are of interest for their fundamental structural and magnetic properties such as weak itinerant magnets when $A$ = Y and La. In this study, the crystal structure and magnetic properties of La$_{2-x}$A'$_x$Ni$_7$ compounds with magnetic rare earth elements ($A'$ = Sm, Gd) have been investigated combining X-ray powder diffraction and magnetic measurements. These intergrowth compounds crystallize in a mixture of 2$H$ hexagonal (Ce$_2$Ni$_7$-type) and 3$R$ rhombohedral (Gd$_2$Co$_7$-type) polymorphic structures which are related to the stacking of [$AB_5$] and [$A_2B_4$] subunits along the $c$-axis. The average cell volume decreases linearly versus $A'$ content, whereas the $c/a$ ratio reaches a minimum at $x = 1$, due to geometric constraints upon $A'$ for La substitution between the two different subunits. The magnetic properties strongly depend on the structure type and the $A'$ content. Hexagonal La$_2$Ni$_7$ is a weak antiferromagnet (wAFM) at low field and temperature and undergoes metamagnetic transitions towards weak ferromagnetic state (wFM) under applied field. Under an applied field of 0.1 T, La$_{2-x}$A'$_x$Ni$_7$ intermetallic compounds display two different transition temperatures $T_1$ and $T_2$ that both increase with $x$. $T_1$ is associated with a wFM-wAFM transition in the 2$H$ phase for $A'$= Sm, whereas $T_2$ is related to the Curie temperature of both 2$H$ and 3$R$ phases. A metamagnetic behaviour is observed between $T_1$ and $T_2$ with transition field $\mu_0H_{Trans}$ between 2 and 3.5 T for compounds with $A'$ = Sm. The La$_{2-x}$Sm$_x$Ni$_7$ compounds ($x > 0$) behave as hard magnets with a large coercive field $\mu_0H_C$ at low temperature ($\mu_0H_C > 9$ T at 5 K for $x = 2$), whereas the La$_{2-x}$Gd$_x$Ni$_7$ compounds ($x > 0$) are soft ferrimagnets with a linear increase of the saturation magnetization versus Gd content.

---

*Corresponding author: valerie.paul-boncour@cnrs.fr, Phone number: +33 1 49 78 12 07

## 1-Introduction

$AB_n$ intermetallic compounds ($A$ = rare earth and $B$ = transition metal, $n$ =1-13) exhibit a wide range of magnetic properties that have been the subject of extensive research, both at the fundamental level and with regard to their potential applications, including permanent magnets, magnetocaloric properties and magnetostrictive materials [1]. Furthermore, numerous studies have been conducted on these compounds with regard to their hydrogenation and electrochemical properties, particularly in relation to their potential use as negative electrodes in Ni-MH batteries [2].

The metamagnetic properties of $La_2Ni_7$ have been the subject of interest for many years [3-13]. Due to its weak itinerant antiferromagnetic properties, it can be considered as a fragile magnet, i.e. a compound where the magnetism (ordering temperature and magnetic moment) can be suppressed by perturbations such as doping, applied pressure or applied field [14]. Such systems are good candidates to find new superconducting materials near quantum critical point. In addition, recent studies of both $La_2Ni_7$ and $Y_2Ni_7$ intermetallic compounds have established a correlation between their crystal structure and their magnetic properties [10, 11]. These two compounds crystallize in two different polymorphic intergrowth structures: $La_2Ni_7$ is hexagonal ($Ce_2Ni_7$-type, $P6_3/mmc$ space group), whereas $Y_2Ni_7$ is rhombohedral ($Gd_2Co_7$-type, $R$-$3m$ space group). These structures result from the stacking of one [$A_2Ni_4$] and two [$ANi_5$] subunits along the $c$-axis repeated twice for the hexagonal $2H$ phase, and three times for the rhombohedral $3R$ phase.

$La_2Ni_7$ with $2H$ structure was first considered as a weak antiferromagnet (wAFM) at low field with Néel temperature, $T_N$ = 55(2) K [4, 8, 10, 12, 13]. $Y_2Ni_7$ with $3R$ structure is a weak ferromagnet (wFM) with Curie temperature, $T_C$ = 53(2) K [15-19]. Their magnetic properties are related to a narrow band at the Fermi level, which is close to the onset of ferromagnetism according to the Stoner criterum [10, 19]. The wAFM order observed in $La_2Ni_7$ has been related to its hexagonal structure, the symmetry of which allows an inversion of the direction of the Ni moments around the Ni atoms located in the [$La_2Ni_4$] subunit. This has been proposed first by ab-initio DFT calculations [10] and then confirmed by neutron diffraction on a single crystal [13]. This last study revealed the existence of three magnetic structures at low field: one commensurate and two incommensurate wAFM structures. The temperature at which the transition occurred was measured at zero field ($T_1$ = 61.0 ±0.2K, $T_2$ = 56.5 ± 0.2K, and $T_3$ = 42.2 ± 0.2K) and found to vary with the applied field [12, 13, 20]. $La_2Ni_7$ undergoes a metamagnetic transition towards a ferromagnetic state with a transition field $\mu_0 H_{trans}$ of 4.6 T at 5 K [9, 11].

The experimental study of the $La_{2-x}Y_xNi_7$ pseudobinary compounds has shown that for $x \leq 1$ the compounds crystallize in the hexagonal structure with Y atoms substituted to La only in the [$A_2Ni_4$] subunit, whereas for $1 < x < 2$ a mixture of $2H$ and $3R$ phases is observed when Y is substituted on both [$A_2Ni_4$] and [$ANi_5$] subunits [11]. The pseudobinary $La_{2-x}Y_xNi_7$ ($x \leq 1$) compounds, with only hexagonal structure, display a wAFM ground state with a Néel temperature increasing with the Y content. Furthermore, a metamagnetic transition from wAFM towards a wFM state is observed at low temperature, with the transition field increasing with increasing Y content [11]. This indicates that the wAFM state becomes more stable than the wFM state upon Y substitution in the 2H structure. For $x > 1$, the samples contain a mixture of 2*H* and 3*R* phases and are wFM with the same $T_C = 55$ K independently of Y content and the phase composition [11].

Several studies have been published for binary $A_2Ni_7$ compounds with magnetic *A* element [3, 5, 21]. Most of them relate to their structural and hydrogenation properties [22-27]. It has been observed that the relative ratio of hexagonal (2*H*) versus rhombohedral (3*R*) structure depends on the size of the *A* elements [28]. $La_2Ni_7$ and $Ce_2Ni_7$, which have the largest *A* atom radius, can be easily synthesized to crystallize only in the hexagonal structure [4, 29]. $A_2Ni_7$ compounds with the smallest *A* radii (*A* = Y, Ho, Dy, Er) are obtained with only the rhombohedral (3*R*) structure [28]. The $A_2Ni_7$ compounds with intermediate *A* radii (*A* = Pr to Tb) generally show a mixture of both 2*H* and 3*R* phases, the ratio of which depends on the radius of the *A* atom [22, 25, 27, 28]. However, only a few studies have been devoted to the magnetic properties of $A_2Ni_7$ compounds. Their Curie temperature $T_C$ increases from 50 K for $Ce_2Ni_7$ [5] to a maximum of 116 K for $Gd_2Ni_7$ [21] and then decreases progressively down to 67 K for $Er_2Ni_7$ [30, 31]. Their $T_C$ are higher than those of $ANi_3$, $ANi_2$ and $ANi_5$ compounds, but lower than those of $A_2Ni_{17}$ compounds [30]. Zhou *et al*. [32] investigated the field dependence of the magnetization at 4.2 K and up to 35 T of $A'_{2-x}Y_xNi_7$ compounds (*A'* = Gd, Tb, Ho) with $x \geq 1.8$. They have used these results to determine the magnetic coupling between the rare earth and Ni $J_{ANi}$ with a mean field model and found them twice larger than for $A_2Ni_{17}$ compounds. Nevertheless, these studies have not been reproduced and no recent work has been published on the magnetic properties of the $A_2Ni_7$ compounds (*A* = magnetic rare earth).

To gain a better understanding of the magnetic properties of $A_2Ni_7$ compounds containing a magnetic rare earth, the study of pseudobinary $La_{2-x}A'_xNi_7$ compounds was undertaken. Non-magnetic La was substituted by either a light (*A'* = Sm) or a heavy (*A'* = Gd) magnetic rare earth, which influence the sign of the magnetic interaction with neighbouring transition metals. Sm and Gd single ions are also characterized by different free ion moments ($\mu_{Sm} = 0.84$ $\mu_B$ and $\mu_{Gd} = 7.94$ $\mu_B$) and anisotropy (elongated ellipsoid for Sm and spherical shape for Gd). Several $La_{2-x}A'_xNi_7$ compounds were synthesized for $0 \leq x \leq 2$ and characterized by X-ray diffraction and magnetic measurements. The replacement of La by Sm or Gd magnetic rare earths is expected to induce ferromagnetic or ferrimagnetic order due to their localized 4*f* moments compared to the weak itinerant magnetism of Ni atoms. In addition, the potential correlation between the polymorphic structure of these compounds and their magnetic properties, as observed in the La and Y systems, prompted this investigation. The results will be compared and discussed in the context of previous literature on the subject.

## 2-Experimental methods

Polycrystalline samples with composition $La_{2-x}A'_xNi_7$ (*A'* = Sm, Gd, $0 \leq x \leq 2$) were prepared by induction melting of the pure elements La (99.9 %), Gd and Sm (Alfa Aesar 99.9%), Ni (Praxair 99.95%) under a purified argon atmosphere. The $La_{2-x}Gd_xNi_7$ alloys were sealed under argon in silica tubes and were annealed for 7 days at 1273 K [27, 33]. As for the $La_{2-x}Sm_xNi_7$ alloys, an excess of Sm was necessary. These ingots were crushed into powder, pressed into pellet, wrapped in Ta foils, sealed under argon in a stainless steel crucible and annealed 3 days at 1223 K [27].

The samples were characterized by X-ray diffraction (XRD) using a D8 diffractometer from Bruker with Cu $K_α$ radiation. The XRD patterns were refined by the Rietveld method using the FullProf code [34]. The XRD patterns have already been published in [11, 33]. The analysis of their chemical composition by electron probe micro-analysis was already reported in [27, 33].

Magnetic measurements were performed using a Physical Properties Measurement System (Quantum Design PPMS-9) operating from 2 to 300 K, with a maximum applied field of 9 T. The measurements were done with few mg of sample (bulk piece or powders fixed with a resin) which were placed in gelatin capsules.

## 3. Results and discussion

### 3.1. Structural properties

In Figure SI1 (Supplemental Information), the percentage of $2H$ and $3R$ phases of binary $A_2Ni_7$ compounds annealed at 1273 K has been plotted versus the $A$ atom radius (see ref [35] and references herein). The data clearly demonstrate the significant influence of geometric constraints induced by the lanthanide contraction on the stability of the polymorphic structure [22, 25, 27, 28]. The following presents the investigation of two solid solutions between $La_2Ni_7$ and $Sm_2Ni_7$ and $Gd_2Ni_7$ binary compounds, which will be denoted $La_{2-x}A'_xNi_7$, where $A'$ represents the specific rare earth element substituted for La.

All the synthesized $La_{2-x}A'_xNi_7$ compounds crystallize in hexagonal ($Ce_2Ni_7$-type, $P6_3/mmc$) and/or rhombohedral ($Gd_2Co_7$-type, $R-3m$) structures as evidenced by their XRD patterns published in our previous work [33]. In these structures, the $A$ atoms occupy two different Wyckoff sites belonging either to $[A_2Ni_4]$ or to $[ANi_5]$ subunits, while the Ni atoms are in 5 different Wyckoff sites (one in the $[A_2Ni_4]$ subunit, two in the $[ANi_5]$ subunit, and two at the interfaces between the two subunits) as detailed in ref. [11]. The results presented here are for $A'$ = Sm or Gd, but the results previously obtained for $A'$ = Y in ref. [11] have been added to the figures 1 and 2 to highlight the influence of the $A'$ atom size on the cell parameter variation. The cell parameters of the $La_{2-x}A'_xNi_7$ compounds ($A'$ = Y, Sm and Gd) in both $2H$ and $3R$ structures are shown in Figure 1 and the weight percentage of the hexagonal phase in Figure 2. The $c$ parameter of the $3R$ phase is multiplied by 2/3 to compare them with the hexagonal phase, while the cell volume is expressed in formula units. In all cases, the $a$, $c$ and $V$ parameters decrease with the $A'$ content due to their smaller radii compared to that of La. No significant difference in relative $a$ and $c$ cell parameters is observed between $2H$ and $3R$ phases. The influence of the size of the $A'$ element is clearly observed in the variation of the relative cell parameters (La > Sm > Gd > Y) whatever the polymorph structure. The relative cell volumes all decrease linearly with the rate of substitution following the Vegard law expected for a solid solution. The d$V$/d$x$ slopes are -4.27(9), -3.83(7) and -3.13(4) Å$^3$/f.u./$x$ for $A'$ = Y, Gd and Sm, respectively.

In all systems and regardless of the structure, $c/a$ shows a minimum for $x = 1$ that can be explained by the preferential substitution of La by a smaller $A'$ element in the [$A_2$Ni$_4$] subunit up to $x = 1$. Then, once all the La atoms of the [$A_2$Ni$_4$] have been substituted by $A'$ atoms, the substitution of $A'$ for La in the [$A$Ni$_5$] subunit starts for $x > 1$ [11]. The cell contraction is more pronounced along the $c$-axis to accommodate the strains induced by the substitution solely in the [$A_2$Ni$_4$] subunit (for $x \leq 1$). Then, for $x > 1$, the contraction occurs in the ($a, b$) plane when La is replaced by $A'$ element in both types of subunits. As the $A'$ radius increases ($r_Y < r_{Gd} < r_{Sm}$) the $c/a$ ratio variation becomes smaller. This shows that the $A'$ substitution affects the anisotropic lattice variation and that the strains are greater when the La and A' atom radii differ more. The samples crystallize in a single hexagonal phase up to $x = 1$ for $A'$ = Y and Gd (Fig. 2) whereas all compounds with $A'$ = Sm contain a mixture of 2$H$ and 3$R$ phases. For $A'$ = Y and Gd and $x > 1$, a mixture of 2$H$ and 3$R$ phases is observed with a progressive reduction of the 2$H$ fraction (Fig. 2). As mentioned above, only Y$_2$Ni$_7$ was obtained in the pure rhombohedral structure.

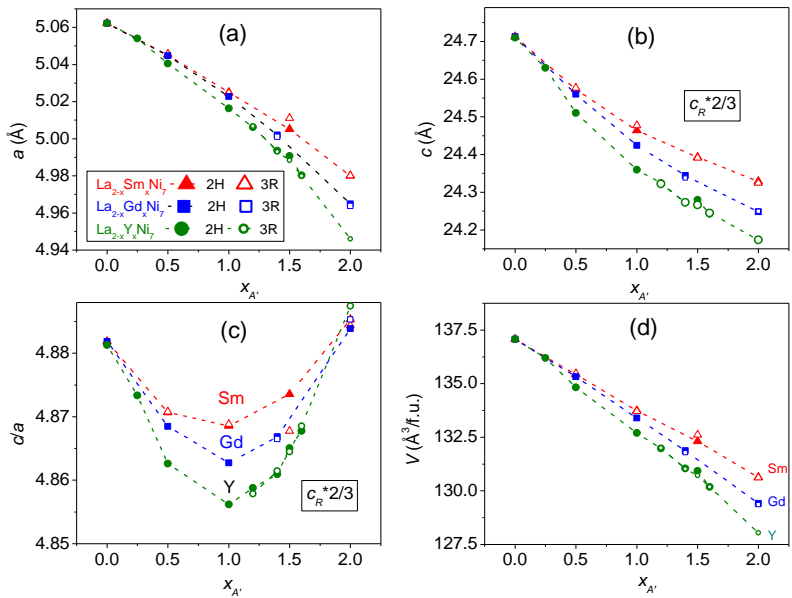

Fig. 1. Cell parameters of La$_{2-x}A'_x$Ni$_7$ compounds: a) $a$ parameter, b) $c$ parameter, c) $c/a$ ratio and d) unit cell volume $V$. To facilitate the comparison between the two polymorphs the $c$ cell parameters have been multiplied by 2/3 for the rhombohedral phase.

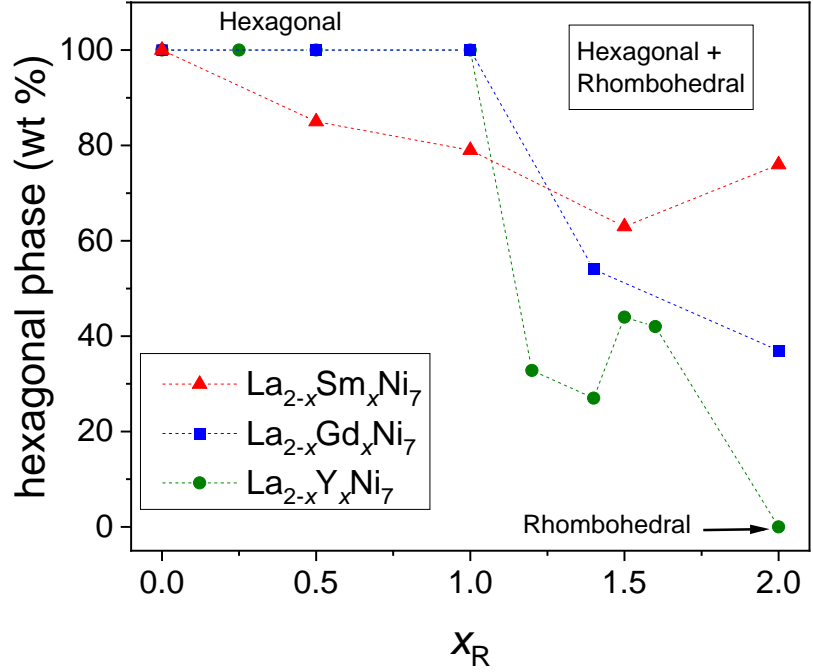

Fig. 2. Weight percent of hexagonal phase (*2H*) determined by XRD pattern Rietveld refinement for La$_{2-x}$A'$_x$Ni$_7$ compounds with *A'* = Sm, Gd or Y. The results for *A'* = Y are from ref. [11] and those of *A'* = Sm and Gd from [33].

### 3.2. Magnetic properties

The magnetic curves *M(T)* of La$_{2-x}$Sm$_x$Ni$_7$ and La$_{2-x}$Gd$_x$Ni$_7$ compounds measured under an applied field of 0.1 T are compared in Fig. 3a and b respectively. Several *M(T)* curves (for $x > 0$) indicate the existence of two distinct ordering transition temperatures, hereafter denoted $T_1$ and $T_2$, with $T_1 < T_2$. The evolution of the transition temperatures (Fig. 4a and b) demonstrates that both $T_1$ and $T_2$ increase with the *A'* content. The Néel temperature $T_N$ of La$_2$Ni$_7$ is between $T_1$ and $T_2$ for $x_{Sm} \leq 1$ and for $x_{Gd} \leq 2$. The observed increase in temperature versus *A'* content provides evidence that these variables are related to the influence of magnetic rare earths on magnetization. In order to gain insight into the origin of these two transition temperatures, the magnetic behaviour of the Sm$_2$Ni$_7$ and Gd$_2$Ni$_7$ binary compounds will be first investigated. This will be followed by an examination of the solid solutions.

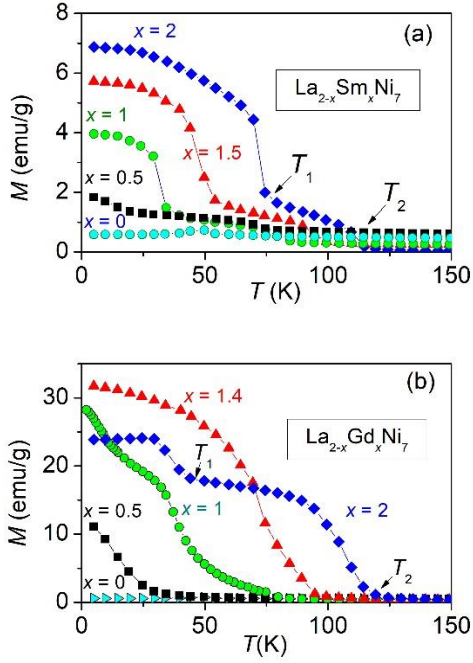

Fig. 3. $M(T)$ curves of a) $La_{2-x}Sm_xNi_7$ and b) $La_{2-x}Gd_xNi_7$ compounds measured under an applied field of 0.1 T in field-cool (FC) mode. The $T_1$ and $T_2$ temperatures are indicated by arrows for the $Sm_2Ni_7$ and $Gd_2Ni_7$ binary compounds.

The two different ordering transition temperatures $T_1$ and $T_2$ are reported in Table 1 for the $A'_2Ni_7$ binary compounds. Both $Sm_2Ni_7$ and $Gd_2Ni_7$ compounds are composed of a mixture of $2H$ and $3R$ phases, which can exhibit distinct magnetic behaviours. It is noteworthy that the relative intensity of magnetization before and after $T_1$ differs markedly between the two compounds. It seems reasonable to posit that this discrepancy is attributable to the relative weight fraction (wt%) of the $2H$ and $3R$ phases. The determination of the magnetic phase weight fractions and the magnetic ordering temperatures $T_1$ and $T_2$ are detailed in supplementary materials (Figs. SI2 and SI3 for $Sm_2Ni_7$ and $Gd_2Ni_7$ respectively). The contribution of the magnetization in the paramagnetic range is first extrapolated linearly to low temperature and then subtracted from that of the magnetically ordered phase. Then two values $\Delta M_1$ and $\Delta M_2$ are obtained from the jump of the magnetization at 5 K and in the middle of the plateau between $T_1$ and $T_2$ respectively. The relative weight fraction $W_{Mag}$ of the magnetic phase between $T_1$ and $T_2$, is then calculated according to:

$$W_{Mag} = 100 \times \frac{\Delta M_2}{\Delta M_1} \qquad (1)$$

It can be assumed that at 5 K, the magnetization of the $2H$ and $3R$ phases as well as their magnetic anisotropy are very close. Consequently, $W_{Mag}$ is compared to the weight fractions of $2H$ and $3R$ phases ($W_{XRD}$) obtained from the refinement of the XRD patterns (Table 1). The close values obtained for $W_{Mag}$ and $W_{XRD}$ of the $3R$ phase suggest that the magnetization between $T_1$ and $T_2$ is predominantly associated with the contribution of the rhombohedral phase, whereas that of the hexagonal phase becomes significantly less pronounced. A similar quantitative analysis of the $M(T)$ curves at 0.1 T was performed for all other samples and compared to $W_{XRD}$ of the 3$R$ phase (Figs. 4c and 4d). A relatively good agreement is observed between these two values for all compounds (except for $x = 1.4$ and $A' = $ Gd).

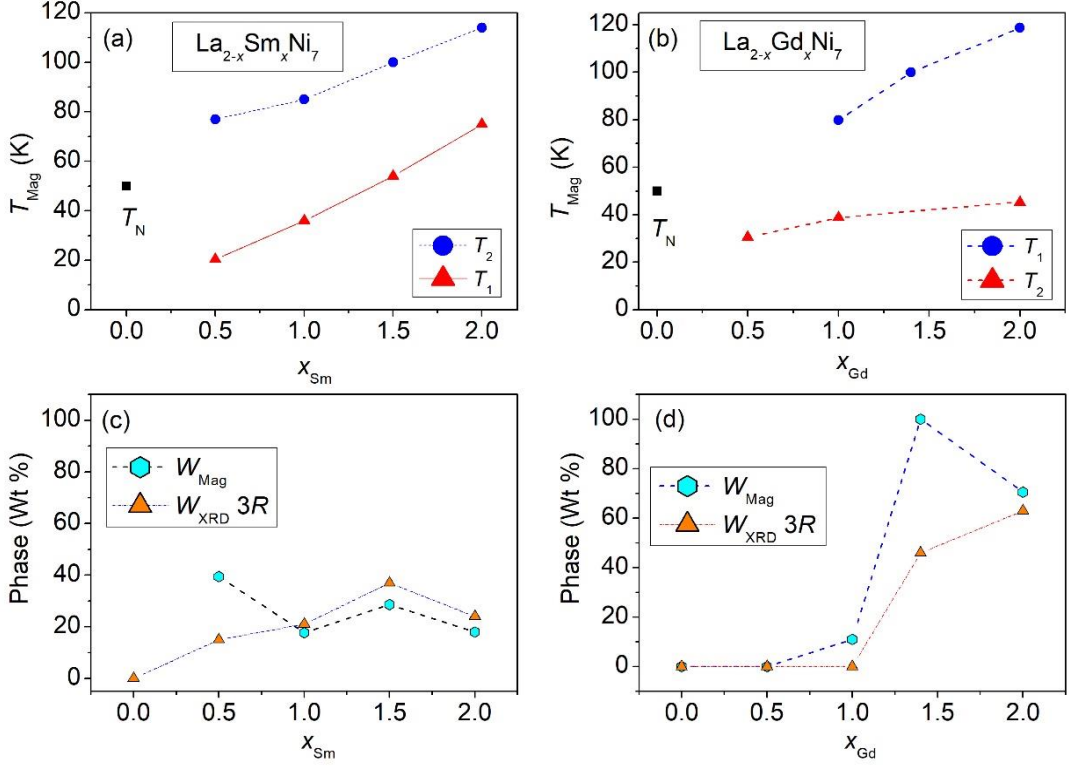

Fig. 4. Magnetic ordering temperatures of La$_{2-x}$A'$_x$Ni$_7$ for (a) $A' = $ Sm and (b) $A' = $ Gd, $W_{XRD}$ of 3$R$ phase and $W_{Mag}$ from magnetization measurements for (c) $A' = $ Sm and (d) $A' = $ Gd.

To further understand the origin of both transitions at $T_1$ or $T_2$, the magnetization curves versus applied field have been measured for all compounds. The compounds with $A' = $ Sm will be initially presented, commencing with Sm$_2$Ni$_7$.

**Table 1:** Magnetic transition temperatures ($T_1$ and $T_2$), relative magnetization ($\Delta M_1$ and $\Delta M_2$), magnetic weight percentage between $T_1$ and $T_2$, and weight percentage of 2$H$ and 3$R$ phases measured by XRD in Sm$_2$Ni$_7$ and Gd$_2$Ni$_7$.

| Phase | $T_1$ (K) | $T_2$ (K) | $\Delta M_1$ (emu/g) | $\Delta M_2$ (emu/g) | $W_{Mag}$ (wt %) | 2$H$ $W_{XRD}$ (wt %) | 3$R$ $W_{XRD}$ (wt %) |
|---|---|---|---|---|---|---|---|
| Sm$_2$Ni$_7$ | 75(1) | 114(1) | 6.8(1) | 1.3(1) | 19(1) | 79(1) | 21(1) |
| Gd$_2$Ni$_7$ | 44(1) | 117(1) | 23.4(1) | 16.5(1) | 70(1) | 37(1) | 63(1) |

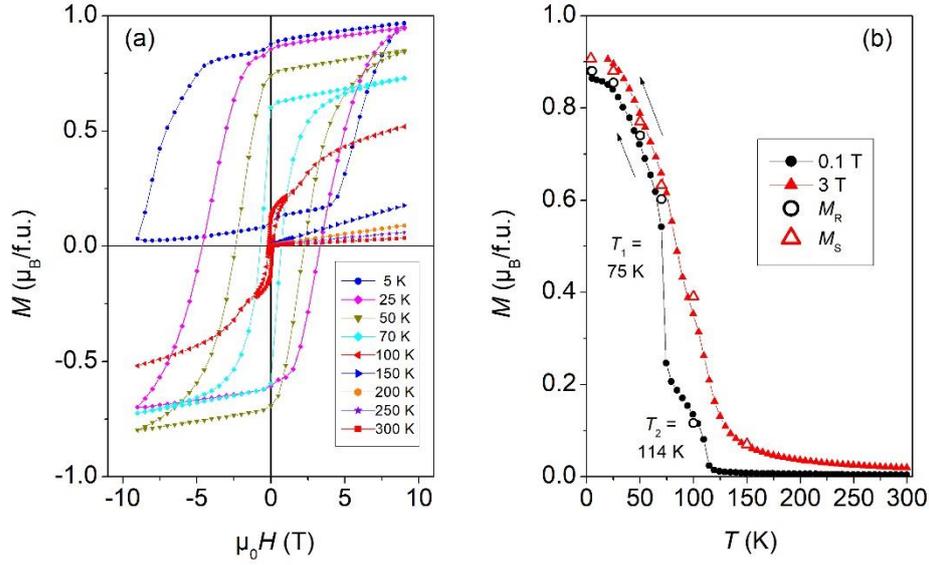

Fig. 5. a) Comparison of the $M(\mu_0 H)$ magnetization curves of $Sm_2Ni_7$ at different temperatures, b) $M(T)$ at 0.1 and 3 T (FC), $M_R$ and $M_S$ obtained from the $M(\mu_0 H)$ curves.

Fig. 5a presents the $M(\mu_0 H)$ curves of $Sm_2Ni_7$ at selected temperatures. Hysteresis loops were measured for temperatures between 5 and 100 K, starting from 9 T and decreasing the applied field down to -9 T, then increasing again to 9 T. At 5 K, only positive values of the magnetization could be measured with a magnetization of 0.033 $\mu_B$/f.u. at -9 T. This means that the coercive field $\mu_0 H_C$ is s larger than the maximum available magnetic field (9T). The extrapolation of $M(\mu_0 H)$ to 0, yields to an estimated coercive field $\mu_0 H_C$ = 9.2(1) T. Upon heating, the coercive field $\mu_0 H_C$ and the remanent magnetization $M_R$ both decrease (Fig. 6a and b). The magnetization curve at 100 K (i.e. at $T_1 < T < T_2$) displays a metamagnetic behaviour with a transition field $\mu_0 H_{Trans}$ of 2 T and a hysteresis loop closing at the same field (Fig. 5a). The S shape of the Arrott plot ($M^2 = f(\mu_0 H/M)$ at 100 K demonstrates the first order character of this transition. The $M(\mu_0 H)$ curves at 150 K and above show a linear behaviour, which is characteristic of a paramagnetic state.

The saturation magnetization ($M_S$) extrapolated from magnetization at high field reaches 0.9 $\mu_B$/f.u. at 5 K. Assuming a collinear ferromagnetic orientation of Sm and Ni moments and a total Ni moment of 0.6 $\mu_B$/f.u. as in $La_2Ni_7$, the calculated Sm moment (0.15 $\mu_B$/Sm) is significantly smaller than that of the free ion value of 0.84 $\mu_B$/Sm. In light of the aforementioned evidence, the diminished $M_S$ value can be ascribed to the presence of a canted magnetic structure, comprising different orientations of the Sm moments in the $[A_2B_4]$ and $[AB_5]$ subunits. This observation corroborates the findings previously documented in the case of rhombohedral $Er_2Ni_7$ by Burzo et al. [31] by neutron diffraction, which have revealed that the two Er sites have different easy magnetization axes, and form a canted ferrimagnetic magnetic structure with a ferromagnetic component along the *c*-axis. It is established that Sm induces a magnetic anisotropy, which becomes particularly pronounced in uniaxial structures.

The existence of the large hysteresis loop at low temperature can be attributed to the energy required to align the Sm moments parallel to the magnetic field. The large coercive field (> 9T) estimated at 5 K can be enhanced by microstructural effects related to the method of preparation, specifically the compression of the powder into pellets prior to annealing treatment used for the samples containing Sm. Adapted heat treatments are currently used to tune the microstructural properties and increase the coercive field of permanent magnets such as $SmCo_5$ [36].

The metamagnetic behaviour observed at 100 K with a transition field $\mu_0 H_{Trans}$ = 2 T can be attributed to a transition from a smaller to a stronger magnetic structure. The $M(T)$ curve measured 3 T (Fig. 5b) demonstrates that the abrupt transition observed at $T_1$ for an applied field of 0.1 T has ceased to exist, and instead, a gradual decline in the magnetization is evident up to $T_2$. The values of the remanent $M_R$ and saturation magnetization $M_S$ have been also reported in Fig. 5b and are found located in the curves at 0.1 and 3 T respectively.

Considering the preceding correlation between $W_{mag}$ and $W_{XRD}$ (3*R*), it can be postulated that the abrupt transition at $T_1$ and for 0.1 T is associated with a transition in the magnetic state of the 2*H* phase, whereby it shifts from a high to a low magnetic state. Meanwhile, the 3*R* phase maintains a high magnetic state and contributes mainly to the magnetization in this temperature range. The low magnetic state of the 2*H* phase above $T_1$, can be either a paramagnetic state or to a wAFM state. If it is the former, then $T_1$ could be identified as the Curie temperature of the hexagonal phase. Alternatively, if it is the latter, then it may be related to a spin reorientation.

In order to elucidate the genesis of this transition and of the metamagnetic phenomenon, it is instructive to draw a parallel with the different magnetic behaviours previously observed in 2*H* $La_2Ni_7$ and 3*R* $Y_2Ni_7$ [10, 11]. Notwithstanding the discrepancy in the magnetic ground state wAFM for 2*H* $La_2Ni_7$ and wFM for 3*R* $Y_2Ni_7$, their magnetic ordering temperatures are remarkably similar, with $T_N$ = 50 K and $T_C$ = 53 K respectively [11]. These transition temperatures were related to the total energy difference between the spin-polarized (magnetic state) and non-spin-polarized (paramagnetic state) for both compounds [10]. Given that the interatomic distances and coordination number remain similar between each Sm and Ni atom in both 2*H* and 3*R* $Sm_2Ni_7$ structures, it is not possible to justify a difference of 39 K in their Curie temperature. Consequently, it seems more probable that there is a spin reorientation of the 2H phase between two magnetic states. Note that a spin reorientation has been previously reported for $Dy_2Ni_7$ and $Ho_2Ni_7$ [37] at a temperature lower of 36 and 40 K than $T_C$ respectively.

In addition, a correlation was established between the crystal structure symmetry and the magnetic phase diagram of 2*H* $La_2Ni_7$ and 3*R* $Y_2Ni_7$ [10]. At 5 K, $La_2Ni_7$ undergoes a metamagnetic transition from a wAFM state towards a wFM state with a transition field of 4.5 T. This has been interpreted, through band structure calculations, by the weak energy difference between the wAFM and wFM states. The wAFM structure, is described by an inversion of the direction of the Ni moments oriented along the $c$-axis below and above a Ni site located in the $[A_2B_4]$ subunit which has a magnetic moment equal to zero. As a sufficient magnetic field is applied, this Ni site recovers a small magnetic moment and the wFM structure with all Ni moment parallel to the c-axis becomes more stable. On the other hand, in the rhombohedral structure, which results from the stacking of three times 1 $[A_2B_4]$ and 2 $[AB_5]$ subunits, such $[A_2B_4]$ Ni site acting as an inversion center is not available. Therefore, 3$R$ $Y_2Ni_7$ has a wFM ground state and no metamagnetic transition is observed. In view of these previous results, it can be assumed that the metamagnetic behaviour observed at 100 K is related to a transition from a wAFM towards a wFM state of the Ni sublattice in the 2$H$ $Sm_2Ni_7$ phase.

In the case of $Sm_2Ni_7$, it is essential to consider the molecular field induced by the Sm moments, which has the potential to act as an external field and stabilize a wFM state for the Ni sublattice of the 2$H$ phase below $T_1$. Upon heating, the Sm moment decreases, which results in a reduction of this molecular field. At $T_1$, this molecular field becomes insufficient to stabilize the wFM structure of the Ni magnetic sublattice belonging to the 2$H$ phase. Consequently, at $T_1$ and in the presence of a low magnetic field, a transition from the wFM to a wAFM state may occur. The application of a transition field of 3 T is sufficient to restore a wFM Ni magnetic sublattice of the 2$H$ phase, as the transition at $T_1$ is no longer observed at such a field (Fig. 5b). The Ni sublattice of the rhombohedral phase remains in a wFM state up to $T_2$, as observed in the case of $Y_2Ni_7$ [10].

The observation of paramagnetic behaviour above $T_2$ indicates that this transition temperature can be attributed to a Curie temperature. As only one transition temperature is observed in $M(\mu_0 H)$ at 3 T, it can be assumed that the 2$H$ and 3$R$ phases, with both a wFM Ni sublattice, have the same Curie temperature. It is also possible for the $T_N$ value of the wAFM state of the 2H phase at 0.1 T to coincide with the $T_C$ of the 3$R$ phase; however, this will be obscured due to the lesser contribution of the former.

The inverse of the magnetic susceptibility in the paramagnetic range ($T > T_2$) can be fitted by a Curie-Weiss law with an effective moment $\mu_{eff} = 1.98(2)$ $\mu_B$/f.u. and a positive paramagnetic Curie temperature $\theta_P = 103(2)$ K, which is slightly lower than $T_2$ (114 K).

In the following, the results of the magnetization curves of the $La_{2-x}Sm_xNi_7$ pseudobinary compounds are presented, considering the results obtained for $Sm_2Ni_7$ and their interpretation. The $M(\mu_0 H)$ curves of $La_{2-x}Sm_xNi_7$ compounds are reported in supplementary materials (Figs. SI5, SI6 and SI7 for $x$= 0.5, 1 and 1.5 respectively).

All the $M(\mu_0 H)$ curves of $La_{2-x}Sm_xNi_7$ ($x \geq 0.5$) compounds display a hysteresis loop at low temperature. The thermal variation of the coercive field $\mu_0 H_C$ and remanent magnetization $M_R$ of the $La_{2-x}Sm_xNi_7$ compounds ($0.5 \leq x \leq 2$) are compared in Fig. 6a and b. The absolute values of $M_R$ are reported for positive (+) and negative magnetizations (-). A discrepancy is evident for $x$= 1, 1.5 and 2 below 35, 50 and 70 K, respectively. This difference is because the maximum available field of 9 T is insufficient to fully saturate the magnetization as the field strength decreases, resulting in incomplete saturation. The values of $\mu_0 H_C$ and $M_R$ decrease as a function of temperature for all compounds, reaching zero above $T_2$. Furthermore, these values decrease in conjunction with the Sm content, supporting the hypothesis that the hysteresis loop depends on the anisotropic behaviour of Sm.

The saturation magnetization $M_S$ at 5 K, slightly decreases between $x = 0$ and 0.5 and, then increases versus $x$ (Fig. 6c). However, all these values are smaller than calculated for a collinear ferromagnetic structure with the free ion value of Sm (0.84 $\mu_B$/Sm) and the total Ni moment measured for $La_2Ni_7$ (7.$\mu_{Ni}$ = 0.6 $\mu_B$) (dashed red curve in Fig. 6c). These lower values can be explained by a canting between the two Sm moments belonging to [$A_2B_4$] and [$AB_5$] subunits as explained for $Sm_2Ni_7$.

The $M(T)$ curves of $La_{1.5}Sm_{0.5}Ni_7$ at different fields from 1 to 9 T show different magnetic transitions *versus* applied field (supplementary materials, Fig. SI4). Some features correspond to a step decrease of the magnetization ($T_1$, $T_2$), whereas the presence bumps with maxima at $T_N$ suggest the existence of AFM states as for $La_2Ni_7$. This reveals a quite complex magnetic phase diagram as observed for $La_2Ni_7$ with probably commensurate and incommensurate AFM and/or canted magnetic structures [12, 13]. The $M(\mu_0 H_C)$ curves present a metamagnetic behaviour from 5 to 60 K (supplementary materials, Fig. SI5). The transition field $\mu_0 H_{Trans}$ of $La_{1.5}Sm_{0.5}Ni_7$ increases to a maximum value of 3.5 T between 40 and 50 K and that of $LaSmNi_7$ reach a maximum of 3.3 T between 50 and 60 K. Concerning $La_{0.5}Sm_{1.5}Ni_7$, the transition field of 3 T is only observed in the curve at 70 K.

These metamagnetic transitions are mainly observed between $T_1$ and $T_2$. The maximum transition field $\mu_0 H_{Trans}$ decreases as a function of Sm content. This reveals the influence of the molecular field induced by Sm on the wAFM-wFM field induced transition of the Ni sublattice. The higher the Sm content, the lower the transition field required to induce this transition. The observed increase of $T_1$ versus Sm content lends support this interpretation. As the Sm content increases, the magnetic ordering temperature of Sm rises, and the molecular field contribution also increases in proportion. For $x \geq 1$, the weight percentage of $3R$ phase derived from $W_{XRD}$ and $W_{Mag}$ are in very good agreement confirming that the $3R$ phase keeps a wFM sublattice up the Curie temperature $T_2$ (Fig. 4a). In addition, the $dM/d\mu_0 H$ slopes obtained from the linear part of the $M(\mu_0 H)$ curves at high field present maxima at temperatures intermediate between $T_1$ and $T_2$ for all $La_{2-x}Sm_xNi_7$ compounds (Fig. 6d). This agrees with a spin reorientation occurring for the $2H$ phase in this temperature range, a larger applied field being necessary to reach a full saturation.

As a paramagnetic behaviour of the $M(\mu_0 H)$ curves is observed above $T_2$, this transition temperature can be associated with a Curie temperature for all pseudobinary compounds and both $2H$ and $3R$ phases. The increase of $T_2$ indicates stronger magnetic interactions induced by Sm.

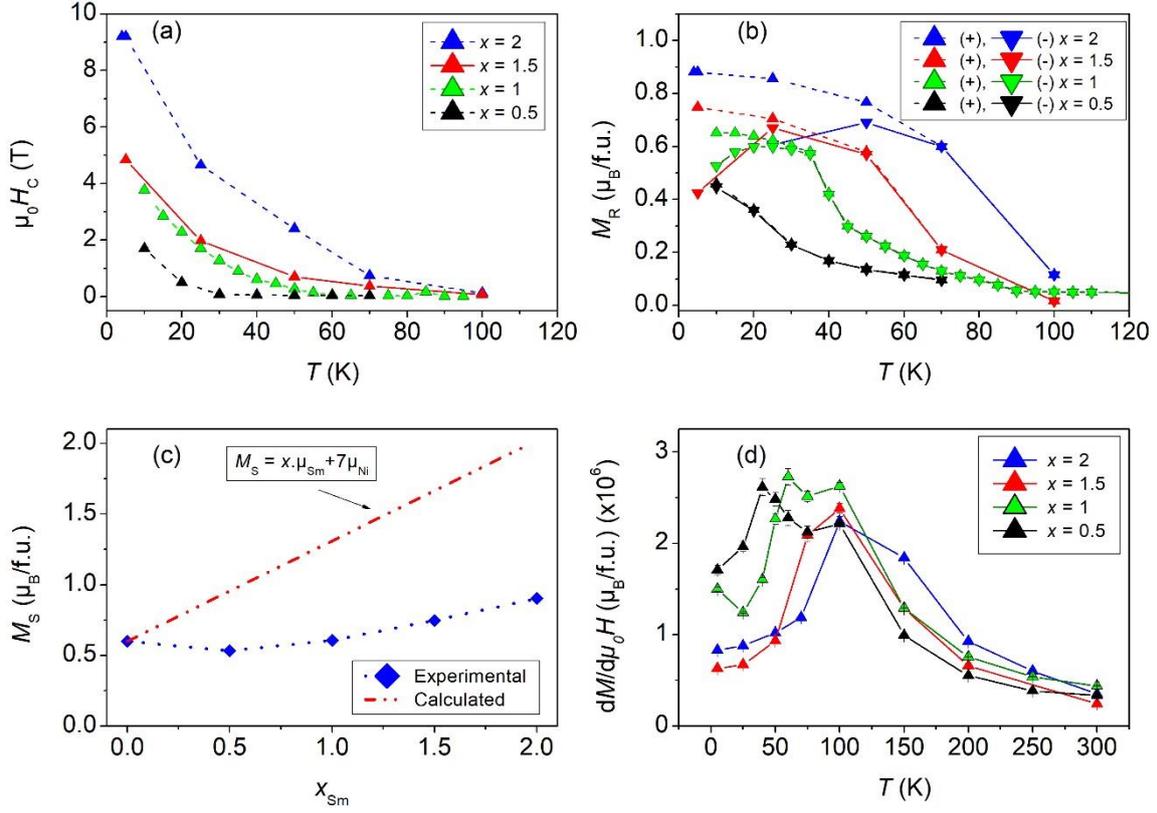

Fig. 6. Evolution of (a) the coercive field $\mu_0 H_C$, (b) the remanent magnetization $M_R$ with (+) = positive magnetization, (-) = negative magnetization (c) the saturation magnetization $M_S$ at 5 K and the calculated total moment assuming a collinear ferromagnetic structure and (d) the slope $dM/d\mu_0 H$ at high field of La$_{2-x}$Sm$_x$Ni$_7$ compounds.

Following an examination of the magnetic properties of compounds containing substitutions of samarium, we will proceed to analyse those of the gadolinium-based compounds. The transition temperature $T_1$ of La$_{2-x}$Gd$_x$Ni$_7$ compounds remains almost constant, whereas $T_2$ increases with the Gd concentration (Fig. 4b). The 3$R$ weight percentage of La$_{2-x}$Gd$_x$Ni$_7$ compounds starts to increase for $x > 1$ (Fig. 4d). However, for $x = 1.4$, as it was not possible to distinguish clearly the two different transition temperatures, this is why $W_{Mag}$ is plotted equal to 100 %.

The $M(\mu_0 H)$ curves of Gd$_2$Ni$_7$ at different temperatures are presented in Fig. 7a and the hysteresis loop at 4 K in Fig. 7b. At 4 K, no coercive field is observed, and the saturation is reached above 1 T. The magnetization decreases upon heating and the $dM/d\mu_0 H$ slope becomes larger. A paramagnetic behaviour is observed above 150 K, i.e. above the Curie temperature ($T_2 = 117$ K). The weak ferromagnetic contribution observed in some of the $M(\mu_0 H)$ curves above 150 K is most probably related to Ni particles segregated at the surface due to the partial oxidation of the rare earth when the samples were crushed into powder.

The saturation magnetization at 4 K, $M_S = 13(1)$ μ$_B$/f.u., is in good agreement with that of Zhou et al. [32], who found $M_S = 12.8$ μ$_B$/f.u. from $M(\mu_0H)$ curves measured at 4.2 K and for field up to 38 T. Gd$_2$Ni$_7$ presents also two transition temperatures, $T_1$ and $T_2$, at 0.1 T (Fig. 7c) and the relative magnetic fraction $W_{Mag}$ is also close to $W_{XRD}$ of the 3$R$ phase as already reported. However, the $M(\mu_0H)$ curves of Gd$_2$Ni$_7$ do not display a metamagnetic behaviour between $T_1$ and $T_2$. A sharp decrease of the initial slope $dM/d\mu_0H$ from 32(2) to 28(2) μ$_B$/f.u/T is observed below and above $T_1$ respectively (inset of Fig. 7c). A second sharp decrease of $dM/d\mu_0H$ is observed at $T_2$.

The evolution of the $M(T)$ curves for applied fields between 0.05 and 0.7 T is displayed in Fig. 7c. The transition observed at $T_1$ for $\mu_0H = 0.05$ to 0.3 T is progressively smeared out at larger field. To explain the decrease of the $dM/d\mu_0H$ slope above $T_1$ it is possible to assume a change of the magnetic structure of the hexagonal fraction above the transition. Nevertheless, with only this small variation of the $dM/d\mu_0H$ slope it remains difficult to identify more clearly the change of magnetic order for the 2$H$ phase. Note that a small bump is observed at 60 K for $\mu_0H = 0.5$ and 0.7 T upon heating, but not upon cooling. As the $M(\mu_0H)$ curves measured at 50 and 60 K are crossing at 0.6 T (Fig. 7a), it can indicate another type of magnetic transition.

The magnetization curves $M(\mu_0H)$ of the La$_{2-x}$Gd$_x$Ni$_7$ compounds are reported in Figures SI8, Si9 and SI10 (supplementary materials) for $x$ = 0.5, 1 and 1.4 respectively. They are characteristic of soft ferrimagnetic materials below $T_2$ and paramagnetic above this temperature, confirming that $T_2$ corresponds to a Curie temperature for all compounds.

The evolution of $M_S$ at 5 K in in La$_{2-x}$Gd$_x$Ni$_7$ compounds versus the rate of Gd (Fig. 8) exhibits a linear increase between $x$ = 0.5 and 2. A linear fit to the data is given by:

$$M_S = x \cdot \mu_{Gd} - 7\, \mu_{Ni} \qquad (2)$$

With μ$_{Gd}$ = 6.73(5) μ$_B$/Gd and μ$_{Ni}$ = 0.057(1) μ$_B$/Ni.

Assuming a collinear ferrimagnetic structure, the measured Gd moment is close to that of the free ion moment (μ$_{Gd}$ = 7 μ$_B$/Gd), whereas the mean Ni moment is slightly smaller than measured for ferromagnetic 3$R$-Y$_2$Ni$_7$ with μ$_{Ni}$ = 0.08(1) μ$_B$/Ni [10, 11]. This behaviour can be compared with the results of Zhou *et al*. [32], who investigated the pseudobinary Y$_{2-x}$Gd$_x$Ni$_7$ compounds for low Gd content ($x \leq 0.2$). In both our work and that of Zhou et al., the Gd is diluted in a non-magnetic rare earth (i.e. La or Y). The exchange-coupling constants $J_{GdNi}/k$ obtained from the magnetization curves at 4.2 K by Zhou et al. [32] was changing from -15.1 K for $x = 0.08$ to -10.8 K for $x = 0.15$ and then to -15.4 K for $x = 0.2$. The shape of the $M(\mu_0H)$ curves for low Gd content ($x = 0.10$ and 0.15) suggested that a field induced ferri-ferromagnetic transition can occur with parallel orientation of Gd and Ni moments at high field. This is no longer the case for $x \geq 0.2$ even for a field up to 38 T [32].

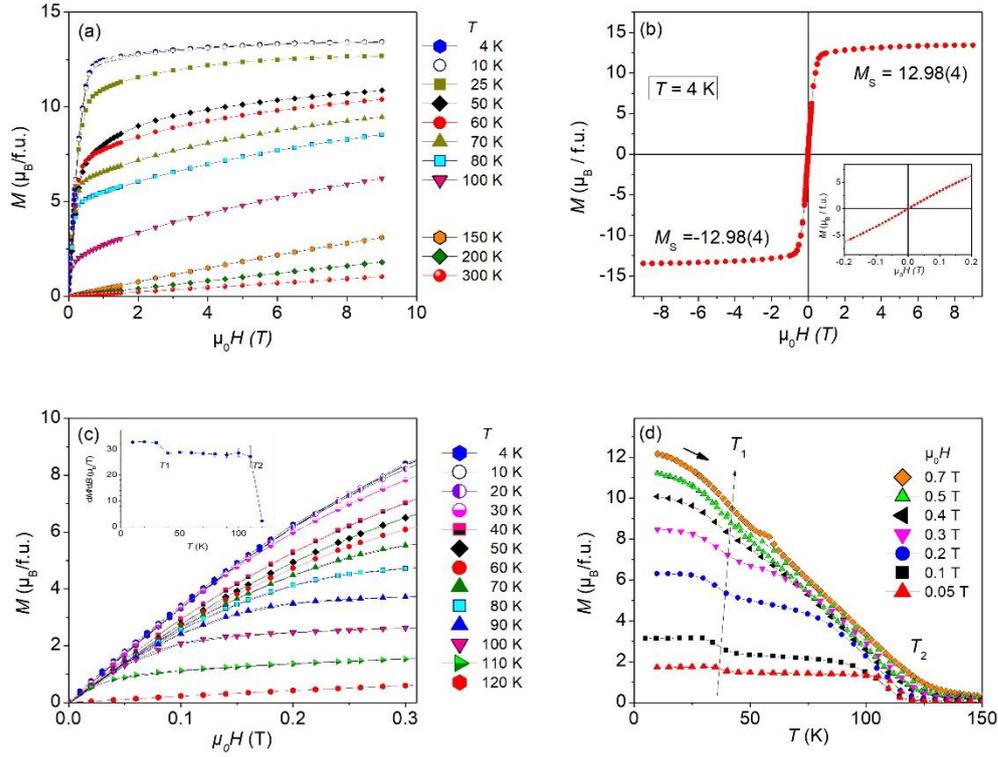

Fig. 7. Comparison of a) the $M(\mu_0H)$ curves at different temperatures, b) $M(\mu_0H)$ at 4.2 K with a zoom of the hysteresis loop in inset, c) $M(\mu_0H)$ curves at low field and initial $dM/dH$ slope in inset d) $M(T)$ curves for several applied fields between 0.05 and 0.7 T of $Gd_2Ni_7$.

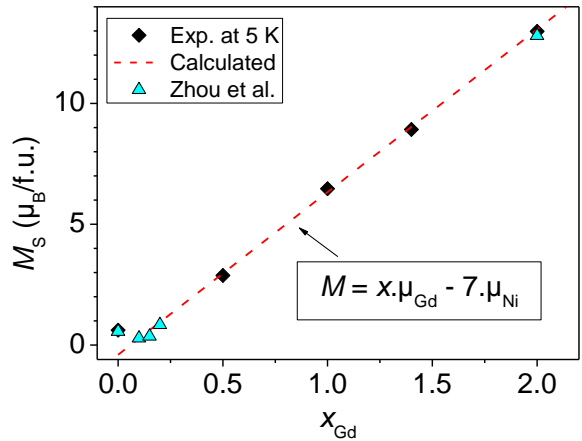

Fig. 8. Evolution of the saturation magnetization $M_S$ versus Gd content at 5 K in $La_{2-x}Gd_xNi_7$ compounds ($x=$ 0 to 2). The dashed red line corresponds to a calculated value of $M_S$ using equation 1. The measurements for $Y_{2-x}Gd_xNi_7$ from Zhou *et al.* [32] were performed at 4.2 K.

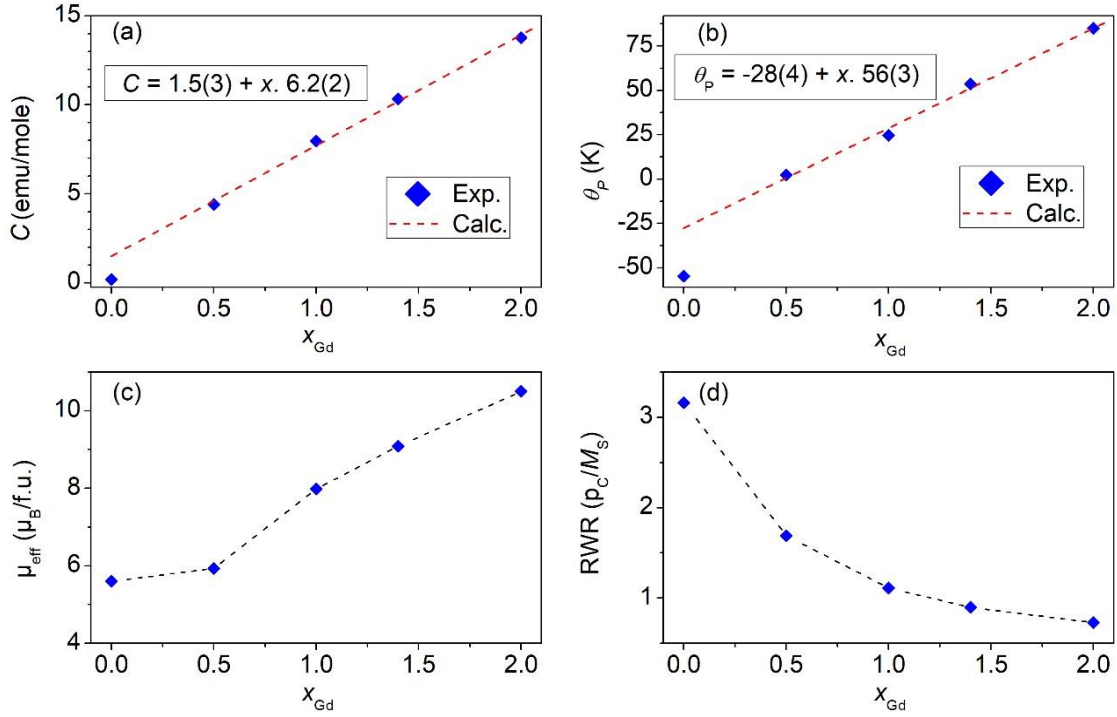

Fig. 9. Evolution of a) the Curie constant and b) the paramagnetic temperature versus Gd content in $La_{2-x}Gd_xNi_7$ compounds ($x = 0$ to 2), c) the effective moment, d) the RWR factor. The data for $La_2Ni_7$ are from ref. [12].

Analysis of the magnetization curves in the paramagnetic region (150-300 K) shows a weak spontaneous magnetization which can be related to the presence of segregated Ni particles on the surface of the grains. This can occur when the samples are crushed to powder. The slopes of these curves were used to obtain the paramagnetic susceptibility $\chi(T)$ of each compound and to calculate the paramagnetic Curie temperature $\theta_P$ and the Curie constant C. These values are plotted in Fig. 9 as a function of Gd content and display a linear variation. The values added for $La_2Ni_7$ are from ref. [12], they were obtained from the magnetic measurements of a single crystal. Both $\theta_P$ and $C$ ($x > 0$) can be fitted with a linear equation, the values of which are plotted in fig. 9. Extrapolation of the calculated curves to $x = 0$ gives larger values than reported for $La_2Ni_7$. This may be due to the influence of the Gd sublattice on the Ni sublattice. It can be observed that $\theta_P$ is significantly lower than the Curie temperature $T_2$ (30 to 50 K lower) and that extrapolation to $x = 0$ gives a negative value. This reflects the competition between ferromagnetic and antiferromagnetic interactions between the Gd and Ni magnetic sublattices. The effective moments are presented in Fig. 9c. For $Gd_2Ni_7$ $\mu_{eff} = 10.5(1)$ $\mu_B$ is smaller than the saturation magnetization ($M_S = 13.1$ $\mu_B$/f.u. at 5 K). The Rhodes–Wohlfarth ratio (RWR = $Pc/M_S$) with $\mu_{eff}^2 = p_c(p_c+2)$ which characterize the itinerant character of a ferromagnet is plotted in Fig. 9d. The value of RWR= 0.73 for $Gd_2Ni_7$ is lower than 1, indicating the localized character of the magnetic transition. This value is comparable with the values obtained for $Dy_2Ni_7$ and $Ho_2Ni_7$ (0.76 and 0.88 respectively) [37] and smaller than for $La_2Ni_7$ (3.2) [11]. As the Gd content decreases the RWR factor progressively increases, reaching 1.7 for $x = 0.5$ (Fig. 9d). This evolution reveals a change from itinerant towards localized character upon Gd for La substitution.

The present study has revealed that the magnetic properties of the $La_{2-x}Sm_xNi_7$ and $La_{2-x}Gd_xNi_7$ pseudobinary compounds are influenced by the difference in crystal symmetry between the two $2H$ and $3R$ structures. This finding is consistent with observations made on $La_{2-x}Y_xNi_7$ compounds [10, 11]. The contribution of the magnetic rare earth $A'$ substituted for La should be considered to explain the evolution of the magnetic properties. The analysis of the magnetization curves has shown that it is possible to discriminate the behaviour of the hexagonal and rhombohedral magnetic phases by the existence of a low and sharp magnetic transition at $T_1$ for a field around 0.1 T. Below $T_1$, the $2H$ and $3R$ phases both order in a canted magnetic structure for $A' = $ Sm or a collinear ferrimagnetic structure for $A' = $ Gd as expected for a heavy rare earth. The influence of the nature rare earth element is also visible at low temperature: for Sm substituted compounds the large hysteresis curves are related to the influence of Sm anisotropy in an uniaxial structure, whereas for Gd, which presents a spherical electronic cloud (corresponding to $\alpha_J = 0$, $\alpha_J$ being the second order Stevens coefficient [38]) no hysteresis is observed.

On heating, the rhombohedral phase retains the same magnetic state with a progressive decrease in the $A'$ and Ni moments up to the Curie temperature $T_2$. For Sm compounds, the hexagonal phase undergoes a magnetic transition at $T_1$ from a wFM to a wAFM Ni magnetic sublattice due to the decrease of the molecular field induced by the Sm moments. Between $T_1$ and $T_2$ the metamagnetic behaviour look like that of hexagonal $La_2Ni_7$. In Gd compounds, the transition at $T_1$ corresponds to a small decrease of the initial slope $dM/d\mu_0H$ of the magnetization curves (Fig. 7c).

The variation of the transition temperatures $T_1$ and $T_2$ versus the $A'$ concentration presents for both systems some similarities. The transition temperature $T_1$ of $La_{1.5}A'_{0.5}Ni_7$ is smaller than $T_N = 55$ K in $La_2Ni_7$ (with $T_1 = 20.4$ K for $A' = $ Sm and 30.5 K for $A' = $ Gd). As the Sm content increases, $T_1$ increases noticeably until it reaches 75 K for the binary $Sm_2Ni_7$. In the case of Gd, $T_1$ also increase with the Gd content, but more moderately, saturating at around 44 K for $Gd_2Ni_7$. The Curie temperature, $T_2$, increases as a function of $A'$ content for both systems reaching 114 and 117 K for $Sm_2Ni_7$ and $Gd_2Ni_7$ respectively. The difference of Curie temperature between the two compounds remains small.

**Conclusions**

$La_{2-x}A'_xNi_7$ ($A' = $ Sm, Gd) pseudobinary compounds crystallize in a mixture of hexagonal (Ce$_2$Ni$_7$-type, $P6_3/mmc$) and rhombohedral (Gd$_2$Co$_7$-type, $R$-$3m$) phases, both constituted by the stacking of two [$AB_5$] and one [$A_2B_4$] subunits along the $c$-axis repeated two and three time respectively. Their cell volumes decrease linearly upon $A'$ for La substitution due to the smaller atomic radius of $A'$. On the other hand, the decrease in $c/a$ ratio exhibits a minimum for $x = 1$. At low field, the magnetization versus temperature curves of $La_{2-x}Sm_xNi_7$ and $La_{2-x}Gd_xNi_7$ show two transition temperatures $T_1$ and $T_2$. $T_1$ has been attributed to a magnetic transition of the hexagonal phase from a high to a lower magnetic state through a spin reorientation of the Ni sublattice, whereas $T_2$ corresponds to the Curie temperature. For Sm compounds, a metamagnetic behaviour is observed between $T_1$ and $T_2$ comparable to that observed for hexagonal $La_2Ni_7$, which can be therefore attributed to a wAFM-wFM transition. The isotherm magnetization curves below $T_1$ reveal that $La_{2-x}Sm_xNi_7$ compounds display remarkable hard magnetic behaviour with a coercive field larger than 9 T for $Sm_2Ni_7$ at 5 K. A canted magnetic structure with different orientations of the Sm moments is assumed to explain the saturation magnetization lower than expected for a collinear ferromagnetic structure with all moments parallel to the $c$-axis. $La_{2-x}Gd_xNi_7$ compounds behave as soft ferrimagnets and the saturation magnetization at 5 K is well fitted assuming a ferrimagnetic structure with antiparallel Gd and Ni magnetic moments. The transition at $T_1$ is also attributed to the hexagonal phase and is characterized by a discontinuous reduction of the initial slope of the $M(\mu_0H)$ curves. Further first principles calculations will be useful to propose stable magnetic ground states and explain the evolution of the magnetic properties.

**Declaration of competing interest**

The authors declare that they have no known competing financial interests or personal relationships that could have appeared to influence the work reported in this paper.

**Acknowledgements**

The project was funded by CNRS. This work has been supported by the French ANR (Agence Nationale de la Recherche) program PROGELEC under the contract MALHYCE ANR-2011-PRGE-006 01. The authors are also thankful to Michel Latroche which was responsible of the MALHYCE project.

**Declaration of generative AI and AI-assisted technologies in the writing process**
During the preparation of this work the authors used [DEEPL / WRITE] to improve the readability of the work. After using this tool/service, the authors reviewed and edited the content as needed and take(s) full responsibility for the content of the publication.

**Data availability**
Data will be made available on request.

**Supplementary Material**

**Relationship between polymorphic structures and magnetic properties of La$_{2-x}$$A'$$_x$Ni$_7$ compounds ($A'$ = Sm, Gd)**

Valérie Paul-Boncour[1*], Véronique Charbonnier[1,2], Nicolas Madern[1,3], Lotfi Bessais[1], Judith Monnier[1], Junxian Zhang[1]

[1]Univ. Paris-Est Créteil, CNRS, ICMPE, UMR7182, F-94320 Thiais, France
[2]Energy Process Research Institute, National Institute of Advanced Industrial Science and Technology (AIST), Tsukuba West, 16-1 Onogawa, Tsukuba, Ibaraki 305-8569, Japan.
[3] Mines Paris, PSL University, MAT - Centre des Matériaux, CNRS UMR 7633, BP 87, 91003, Evry, France

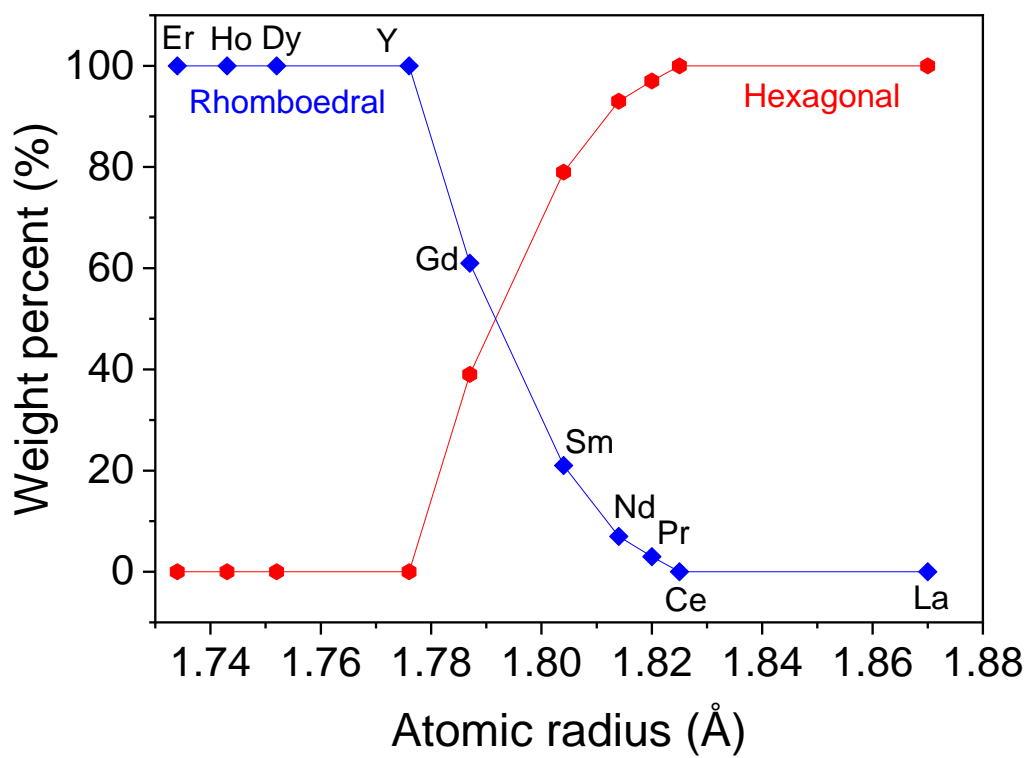

Fig. SI1. Weight percentage of hexagonal and rhombohedral phases of $A_2$Ni$_7$ compounds annealed at 1273 K versus atomic radius. The figure is from V. Charbonnier's PhD thesis [1]. The data are obtained from literature $A$ = La, Ce, Er [2]; Y, Sm, Gd[3] ; Nd[4] ;Pr[5] ;Dy, Ho, [6]

[*]Corresponding author: valerie.paul-boncour@cnrs.fr, Phone number: +33 1 49 78 12 07

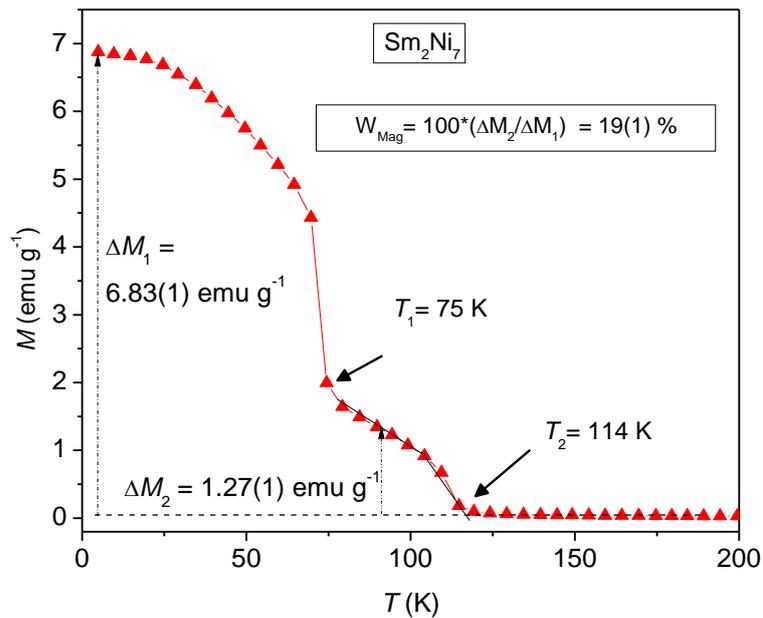

Fig. SI2. Determination of the relative magnetic weight percentages and the transition temperatures $T_1$ and $T_2$, in the binary $Sm_2Ni_7$ compound.

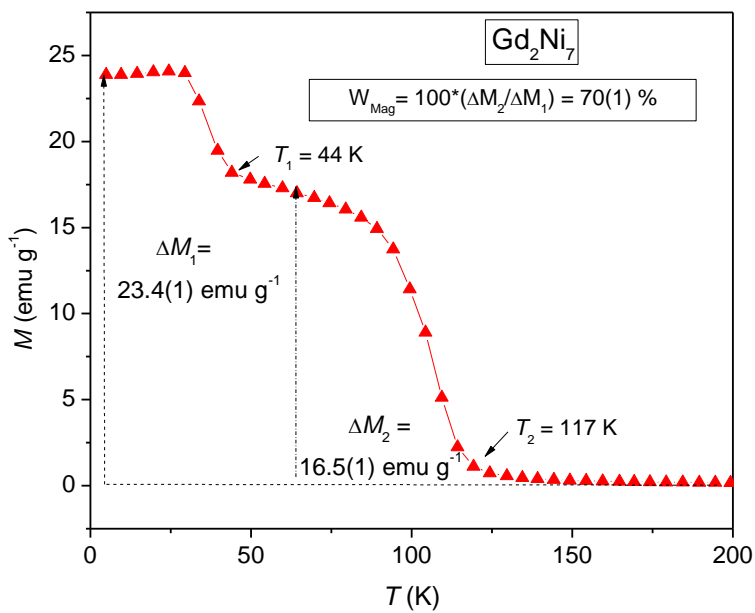

Fig. SI3. Determination of the relative magnetic weight percentages and the transition temperatures $T_1$ and $T_2$, in the binary $Gd_2Ni_7$ compound.

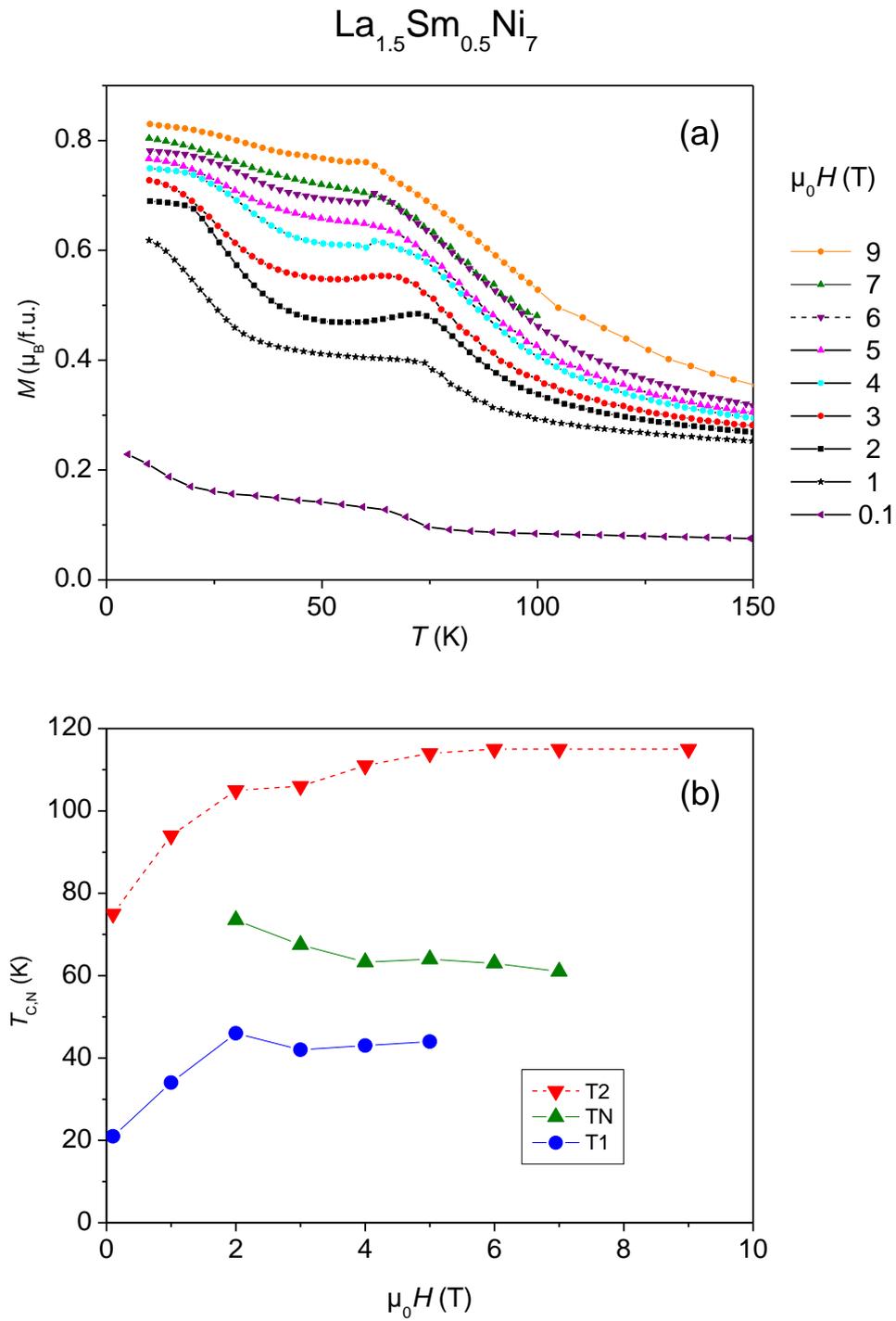

Fig. SI4. (a) Evolution of the $M(T)$ curves of $La_{1.5}Sm_{0.5}Ni_7$ at different applied fields, (b) magnetic transition temperatures as a function of the applied field for $La_{1.5}Sm_{0.5}Ni_7$

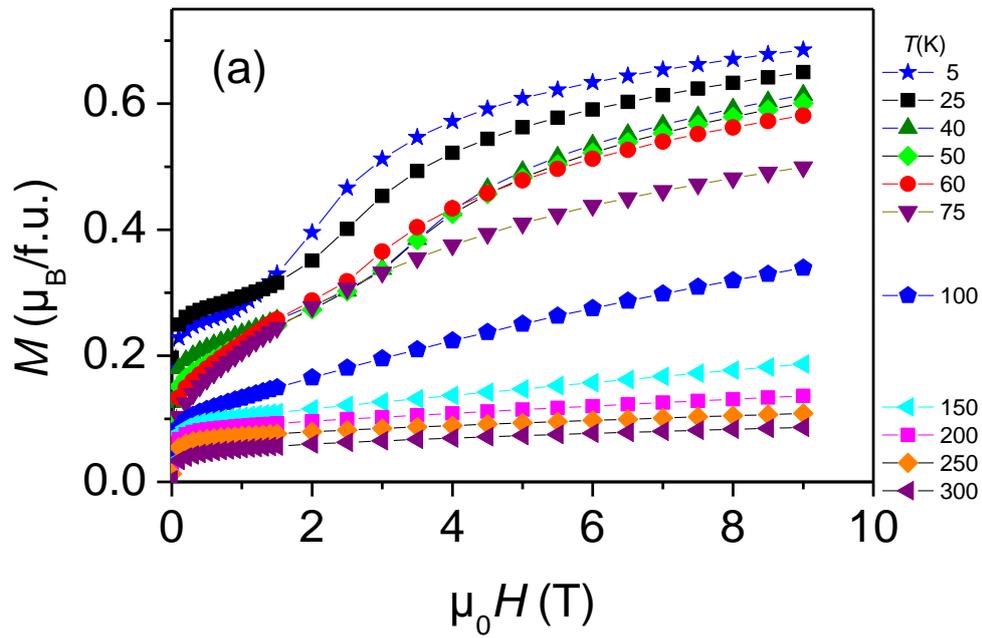

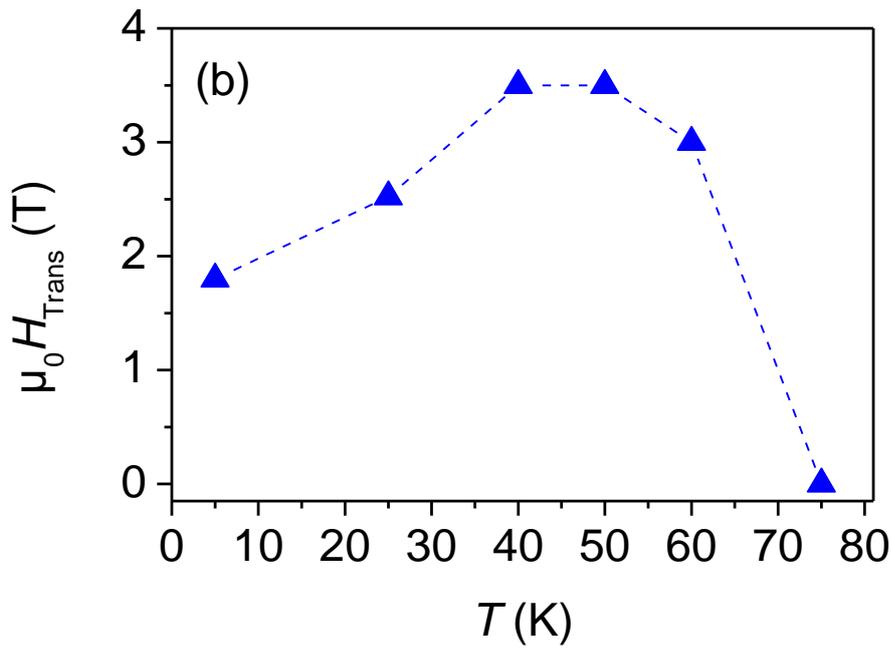

Fig. SI5. (a) Evolution of the $M(\mu_0H)$ curves of $La_{1.5}Sm_{0.5}Ni_7$ at different temperatures. (b) transition field versus temperature for $La_{1.5}Sm_{0.5}Ni_7$.

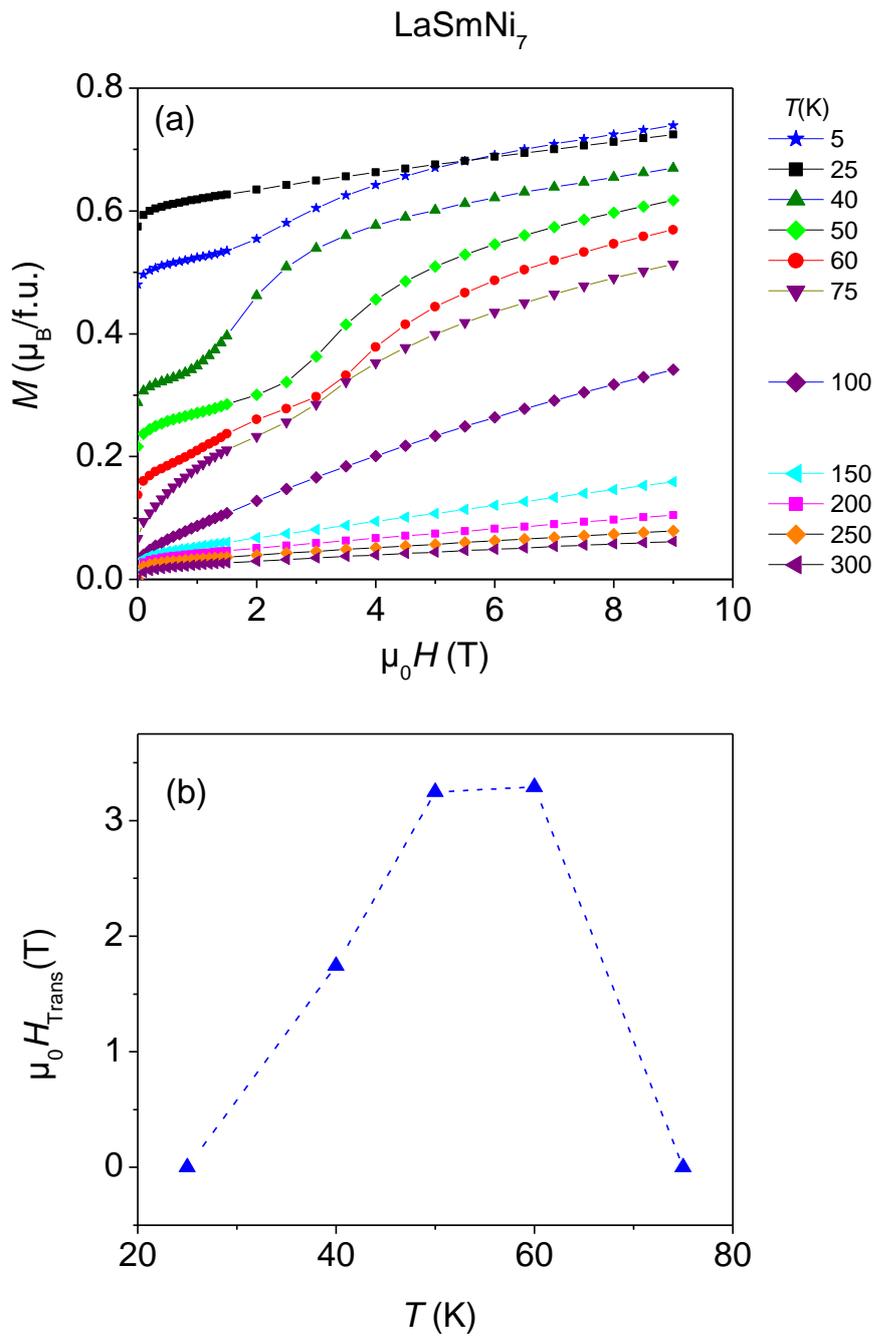

Fig. SI6. Evolution of (a) the $M(\mu_0 H)$ curves of LaSmNi$_7$ at different temperatures, (b) transition field versus temperature.

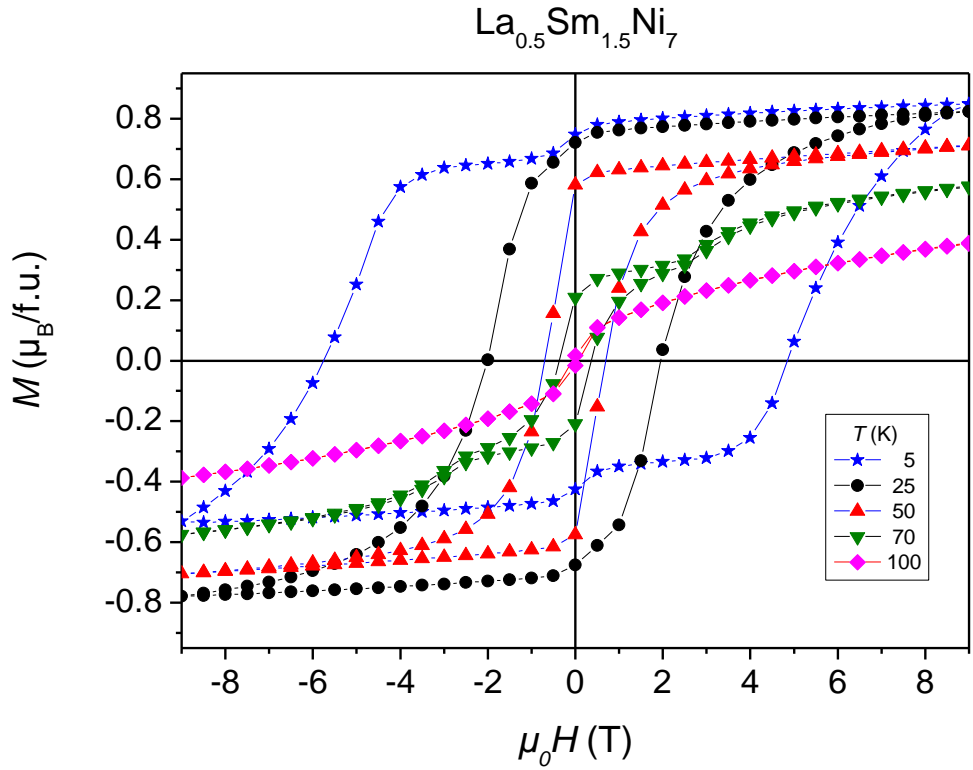

Figure SI7: Evolution of the $M(\mu_0 H)$ curves of $La_{0.5}Sm_{1.5}Ni_7$ at different temperatures.

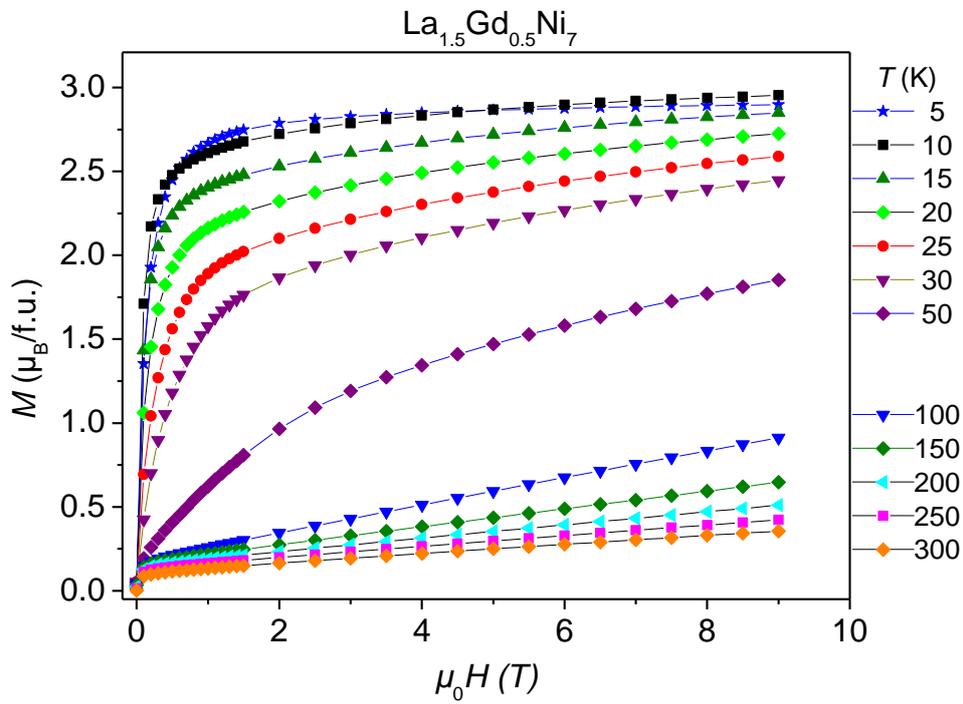

Fig. SI8. Evolution of the $M(\mu_0 H)$ curves of $La_{1.5}Gd_{0.5}Ni_7$ at different temperatures.

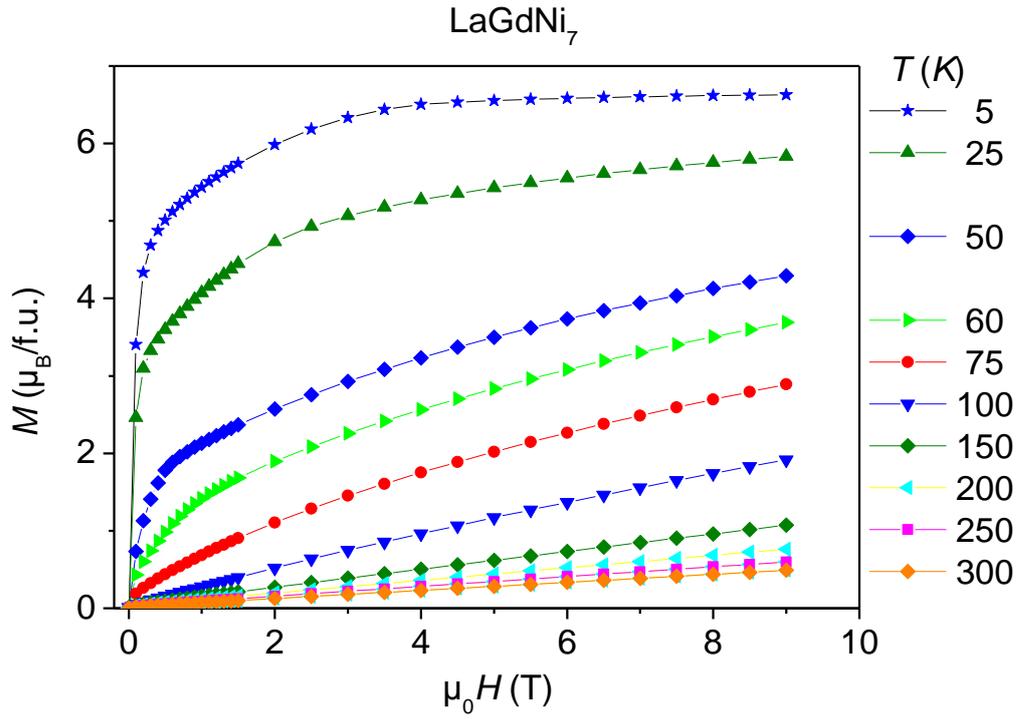

Fig. SI9. Evolution of the $M(\mu_0 H)$ curves of LaGdNi$_7$ at different temperatures.

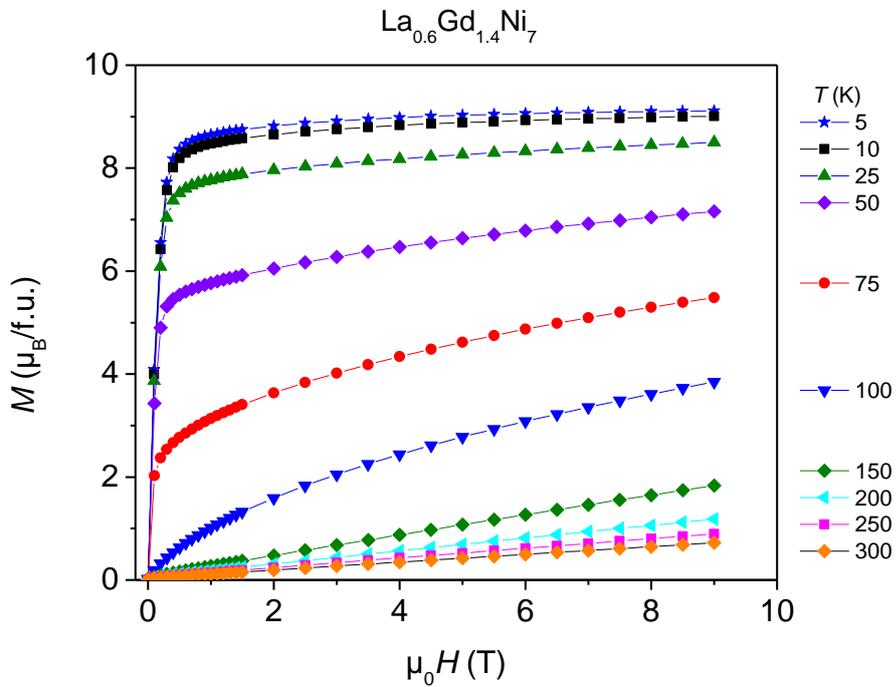

Fig. SI10. Evolution of the $M(\mu_0 H)$ curves of La$_{0.6}$Gd$_{1.4}$Ni$_7$ at different temperatures.